\documentclass[preprint,12pt]{aastex}
\usepackage{epsfig}
\usepackage[dvips]{color}
\usepackage{natbib}
\usepackage{hyperref}
\def\rd{Di\thinspace Stefano}

\slugcomment{Submitted to the Astrophysical Journal}

\shorttitle{PLAN-IT}
\shortauthors{Di Stefano, Matthews, L\'epine } 

\def\zrl{zone for resonant lensing}

\begin{document}
\bibliographystyle{natbib}

\title{Searching for Planets During Predicted Mesolensing Events:\\  
II. PLAN-IT: An Observing Program and its Application to VB~10  
}
\author{Rosanne Di\thinspace~Stefano\altaffilmark{1}, James
  Matthews\altaffilmark{1,2}, and S\'ebastien L\'epine\altaffilmark{3,4}}
\altaffiltext{1}{Harvard-Smithsonian Center for Astrophysics, 60 Garden
  Street, Cambridge, MA 02138}
\altaffiltext{2}{School of Physics \& Astronomy, University of Southampton,
  Southampton, SO17 1BJ, UK}
\altaffiltext{3}{Department of Astrophysics, Division of Physical Sciences,
  American Museum of Natural History, Central Park West 79th Street,
  New York, NY 10024}
\altaffiltext{4}{City University of New York, New York, NY}

%
%

\begin{abstract}
The successful prediction of lensing events
 is a new and exciting
enterprise that 
provides opportunities to discover and study planetary systems.
The companion paper investigates the underlying theory.
This paper is devoted to outlining the components of 
observing programs that can discover planets orbiting stars
predicted to make a close approach to a background star. 
If the time and distance of closest approach can be well predicted,
then the system can be
targeted for individual study. In most cases, however, the predictions
will be imprecise, yielding only a set of probable paths of approach and
event times. We must monitor an ensemble of such systems to ensure discovery,
a strategy possible with observing programs similar to a number of current surveys,  
including PTF and Pan-STARRS; nova searches, 
including those conducted by amateurs;
ongoing lensing programs such as MOA and OGLE;
as well as MEarth, {\sl Kepler} and other transit studies.  
If well designed, the monitoring programs will be guaranteed to
either discover planets in orbits with semi-major axes smaller than
about two Einstein radii, 
or else to rule out their presence. 
Planets on wider orbits may not all be discovered, but if they are common, will
be found among the events generated by ensembles of potential lenses.  
We consider the implications for VB~10, the first star to make a predicted
approach to a background star that is close enough to allow 
planets to be discovered.
VB~10 is not an ideal case, but it is well worth studying. 
A more concise summary of this work, 
and information for observers can be found at 
https://www.cfa.harvard.edu/$\sim$jmatthews/vb10.html.
\end{abstract}

\section{Introduction}

\subsection{A Brief History of Prediction}

When Einstein was first convinced to publish a paper on the generation of
gravitational lensing events, 
he expressed doubt that
the effect would be observed (Einstein 1936). 
The probability of an event occurring
is very low. In addition, detection can be made challenging through the 
 ``dazzling by the light of the much nearer star'',  
the lens.   
Einstein's doubts have not deterred astronomers,
with their familiarity of the universe of possible
lenses and background sources, from predicting
specific lensing events. 

In 1966, Feibelman suggested that   
the star 40~Eridani-a, a member of a triple which also
contains a white dwarf, would pass close enough to a background star
in 1988 to produce detectable lensing effects. Given its mass of 
$0.84\, M_\odot$ and its distance of only $5$~pc, 40~Eridani-a has an
Einstein angle, $\theta_E$, of about $37$~milliarcseconds (mas) when it lenses 
a star at much larger distances from us. Unfortunately, though,  
studies of the motion during the following $20$~years found that the
distance of closest approach would be approximately $3^{\prime\prime}$,
about $81\, \theta_E$ (Feibelman 1986), too
distant for observations at that time to detect the effects of lensing. 
In addition, the closest approach
between 40~Eridani-a and the background star was predicted to occur in June,
when Eridani, a winter constellation, would have been unobservable.  
Any plans to monitor the region were abandoned.

In 2001, Paczy\'nski suggested that the isolated neutron star
RX~J185635-3754 would pass close enough to a background star 
(within about $0.3^{\prime\prime}$)  
that its action as a gravitational lens would produce an astrometric 
shift that could be detected with HST.
As reported by Neuh\"auser et al. 2011, the Ph.D.\, thesis of Eisenbeiss
finds that there is no star close enough to the paths of any 
of the seven isolated neutron stars known as   
the {\sl Magnificent Seven} (Haberl 2007) to produce detectable lensing
through the rest of the 21st century. 

Recently, our group predicted
a very close approach between the high-proper-motion dwarf star VB~10, 
and a dim background star (L\'epine \& \rd\ 2011). With two {\sl HST} images of the region, and
the results of previous of astrometric studies of VB~10, we thought at first
that we could predict 
the distance of closest approach to within several mas, and also
the approximate date of closest approach. Taking into account the
uncertainties, including the uncertainty in 
the motion of the background star, however, 
we realized that existing observations are
consistent with a bundle of paths, each characterized by a  
date and distance of closest approach.
The most likely date was during December of 
2011, when VB~10 is a twilight star, very close to the Sun.
Nevertheless, the predicted approach is very close:
the probability is $\sim 65\%$ that the distance of 
closest approach will lie in the range from less than an Einstein angle to
$15\, \theta_E.$ 
The time of
closest approach could be during the spring of 2012, even though the most
likely dates are earlier. Faced with these uncertainties, 
and also with the fact that the background star is dim,    
we 
set out to assess the value of 
an observing campaign.  
To explore the possibilities in detail, we asked whether, if
VB~10 has planets, they could enhance the lensing signature or even produce 
detectable events at different times. 
The answer to both questions is yes, and the details are
reported in the companion paper (\rd\, et al. 2012a, hereafter Paper I). 

\subsection{Possible Observing Programs} 
 
In spite of the difficulty of predicting individual events, 
Paczy\'nski (1986) found a way to transform the study of lensing
events into
an observational enterprise.
He overcame Einstein's concern about low probability by
pointing out that, fifty years later, computers had made it possible
to track the light curves of many millions of stars per night, guaranteeing the
detection of some events if large, dense stellar fields
could be monitored. He countered the possibility that dazzling by
the lens could prevent the detection of lensing by focusing
on the possibility of lensing by compact dark matter objects.
Within a few years, several monitoring programs were each monitoring
large portions of the Magellanic Clouds and the Galactic bulge to
find evidence of microlensing. The challenge that the first
generation 
 of microlensing observing programs faced was to prove
that they could extract from the haystack of ubiquitous stellar
variability, the signal of a relatively small number of lensing events
(Udalski et al. 1993, Alcock et al. 1993, Aubourg et al. 1993)

They succeeded, and the present generation of lensing monitoring teams
are discovering more than $1500$ events per 
year\footnote{ogle.astrouw.edu.pl/; www.phys.canterbury.ac.nz/moa/}.
By demonstrating that long-term, high-cadence monitoring can
reveal variability of many types, producing a high
science yield, these projects have inspired many other
astronomical time domain studies. These include the
Palomar Transient Factory (PTF; Rau et al. 2009, Law et al. 2009), the 
Panoramic Survey Telescope \& Rapid Response System (Pan-STARRS; Kaiser et al. 2002), 
and the Large Synoptic Survey Telescope (LSST; Tyson 2002).

One particularly innovative approach has been designed to
search for evidence of transits of Earth-like planets around M~dwarfs.
The MEarth Project is a survey to photometrically monitor
approximately 2000 nearby
M dwarfs, searching for transits by planets that could be
 habitable Super-Earths (Irwin et al. 2009a).
The project has had some success in the detection of Super-Earths
(Charbonneau et al. 2009) and also in discovering a number of eclipsing binary
systems (Irwin et al. 2009b; Irwin et al. 2010).
The unique feature that we will find potentially important
in monitoring the sites of predicted lensing events (\S5) is that
MEarth monitors many directions per night, because their
target stars are
spread over a large portion of the sky. The same will be true for
predicted lensing events. 

\subsection{Plan of the Paper}      
We develop an observational strategy to systematically study
predicted lensing events with the goal of learning about the planetary
system of the lens star. The companion paper (\rd\ et al. 2012) presents
the theory. Here we focus on observations. In Sections 2 and 3 we 
describe what we can learn by monitoring the site of a predicted event
during the months and years prior to and after the event, as well
as during the time of closest approach. Section 2 is devoted to
planning observations for
a specific case meant to represent an ideal situation. Section 3 is
devoted to planning a real observing campaign for the star VB~10, 
which is generating the first predicted event about which we can hope to 
derive observational constraints.
In Section 4 we address the issue of how many events are likely to be
predicted each year. Since the majority of predictions will not be 
precise, with the time and distance of closest approach not known
in detail, we turn in Section 5 to observational programs designed
to monitor the sites of thousands of predicted events.
By monitoring a large enough number of events, we can be guaranteed
to detect the lensing signatures of some foreground stars and their
planetary retinues. 
The prospects for successful observing programs are discussed in Section 6.   
In addition to predicted lensing events, there are at least two connections
to ongoing observing programs: The first is to current surveys for lensing events;
the second connection is to transit studies.

\section{Monitoring Individual Predicted Events}

\subsection{Events to target} 

Multiple precise 
measurements of the relative positions of source and lens can, 
in some cases,  allow the
time and distance of closest approach to be computed with modest 
uncertainty limits. If the probability is high that the 
closest approach will lie within $\theta_E$ of the lens, and that
the source star will be observable near the time of 
closest approach,
then monitoring the source star during the lensing event
 can measure the mass of the lens,
determine if it has a planetary system, discover planets in a wide
variety of orbits and with a broad range of masses, and place limits
on the presence of various types of planets (see Paper~I for details) .
In an ideal case, the 
source is brighter than the lens in some waveband, bright enough
that large telescopes are not needed to detect changes of a few percent
 in the amount
of received light. This circumstance can allow amateurs
to play significant roles in the monitoring program, strengthening
the results.      

To make the discussion in this section concrete, we will 
consider the prediction of an event by a specific star,
PMI18537+4929. This is a star with an estimated mass of
$0.46\, M_\odot$. PMI18537+4929 is $44.8$~pc away
and has a proper motion 
$\mu=0.137^{\prime\prime}$~yr$^{-1}$. These parameters
allow us to compute the value of the Einstein angle,
$\theta_E=9.1$~mas. We also have the Einstein-diameter crossing
time, $\tau_E=49$~days. This star falls into the general
category of stars expected to produce events at a rate of just under
one per century.    
  
The Einstein radius, $R_E=\theta_E D_L,$ where $D_L$ is the distance 
to the lens, is $0.41$~AU. If a lensed source is a kpc away, then the
value of the Einstein radius projected onto the source plane is $8.9$~AU.
Thus, even if structures significantly smaller than $\theta_E$ cause
short-lived deviations in the lensing light curve, these features are not likely to 
be washed out by finite-source-size
effects.

The orbital period of a planet with semi-major axis $a=R_E$
is approximately $140$~days. If observations are able to detect deviations
of $1\%$ in the amount of light received from the source star,
then a deviation from baseline is potentially detectable for a 
maximum of $3.5\, \tau_E \sim 172$~days. Thus, rotation could play a role
in increasing the detectability of planets in similar orbits.
Rotation is less likely to be significant in events caused by
wide-orbit planets. On the other hand,
deviations caused by close-orbit planets  
will, in many cases, repeat. (The orbital period for a planet with 
$a=0.25\, R_E$ is about $18$ days.)

For the purposes of the lensing signatures, the important separation is the
projected separation between the lens star and its planet. To simplify
this discussion we consider circular face-on orbits. The value of $a$
is constant, and it we define $\alpha=a/R_E.$ In Paper~I
we have shown that it is convenient to consider
three classes of orbital separation: wide-orbits ($\alpha > 2$); orbits
in the zone for resonant lensing ($0.5 < \alpha < 2$), and close-orbit
planets ($\alpha < 0.5$). 
Events generated by planets
in each of these ranges exhibit distinctive characteristics, suggesting that
the observing strategy should evolve with time, as orbits with different values of 
$\alpha$ become detectable.\footnote{Note that,
although this classification is very useful,
the values for the boundaries given above are
somewhat arbitrary, because event characteristics depend on
the mass ratio $q=m_p/M_\ast,$ where $m_p$ is the mass of the planet
and $M_\ast$ is the mass of the star.}   

An important parameter in determining what we can learn from
a specific predicted event is the angle of closest
approach, $b$, between the foreground and background stars. With
$\beta=b/\theta_E$, three ranges of values determine which
types of planets can be detected. The {\sl wide regime} covers the
entire lens plane; wide-orbit planets can be detected throughout
the wide regime, as long as $\alpha > \beta.$ In addition to wide-orbit
planets, close-orbit planets can be detected if $\beta < \eta,$
where the value of $\eta$ depends on the photometric sensitivity
and is typically larger than $3$ but smaller than $10.$ In addition
to wide-orbit and close-orbit planets, planets in the zone for resonant
lensing can be detected in the {\sl resonant regime,} ($\beta < 2$).
More details are presented in Paper I.   

Because the goal of this section is to determine the best day-by-day monitoring
strategy, we have chosen a star that is more typical than VB~10 of what we can expect
in the future. We do, however, assume that the observing conditions will be better than
those expected for VB~10. First, we assume that the closest approach occurs when the
lens and source are observable: that is, they appear in the night sky, close to neither the Sun 
nor Moon . Second, we assume that the  source star is bright enough that
changes in the light we receive from it of about a percent can be detected.
Third, we assume a distance of closest approach smaller than $\theta_E$. 
We chose a close approach because it allows us to explore the full range of 
behavior which is potentially detectable.

\subsection{Detection as a function of time} 

We considered the possibility that PMI18537+4929 has a Jupiter-mass planet in a face-on, circular
orbit. To study detectability, we conducted a simulation in which 
we generated 20,000 orbits with values of $\alpha$ ranging from $0.1$ to $20.$ For each value
of $\alpha$, we also generated a random starting value for the orbital phase.
We then followed the path of both the star and its planet as the center of mass executes a 
straight-line path making a closest approach of $\beta=0.5$ to a bright background star
at time $t=t_0.$ We computed the magnification, $A_{pl}(t)$, 
caused by the planetary system at
each time, and compared it with the magnification, $A_{pt}(t)$
 expected if PMI18537+4929 alone had been traveling
along the line followed by the planetary system's center of mass. At each time, we asked
whether the value of $A_{pl}(t)-A_{pt}(t)$ had reached an extremum with a value
greater than $0.01.$ If so, we stored the value of $\alpha$ and $(t-t_0)$. 
In order to display the structures in the densest regions, within about $40$~days of $t_0$,
we did not plot all of the points. Nevertheless, the relative density of points in
a given region roughly corresponds to the probability that a planetary system with
a specific value of $\alpha$ will produce a light curve with significant deviations
at time $t.$ 
Points in orange correspond to caustic crossings.  
The magenta curves that start in the upper right and upper left
of the figure were calculated
with the analytic formula given in Equation $3$ of Paper~I, which
applies to planets in wide orbits. The cyan curves that start in the lower right
and lower left 
of the figure were calculated
with the analytic formula given in Equation $6$ of Paper~I, which
applies to close-orbit planets. The black points, from our simulation, 
 generally follow these curves, especially
for the widest and closest orbits; the curves therefore provide good guides
to the positions that would be occupied by low-probability events.

To determine what type of event will be detected as time goes on,
we consider a ruler, with its straight edge
coincident with the vertical axis. As time goes on, the ruler moves
to the right.
The points and/or curves it intersects at each time show the types of events that 
can be detected at that time.

\subsubsection{Early times and late times}

Approximately $500$ days before the closest approach between 
the nearby star and the 
foreground source, a planet with $\alpha=20$ could make a
very close approach to the background star, producing detectable 
lensing effects. 
The radius of the Einstein ring of the planet is
\begin{equation}
\theta_{E,pl}=\sqrt{\frac{m_p}{M_*}}\theta_E=0.42~{\rm mas}~\sqrt{\frac{m_p}{m_J}}. 
\end{equation} 
The Einstein diameter crossing time,
$\tau_{E,pl}$ is approximately $2.3\sqrt{\frac{m_p}{m_J}}$~days. 
If $1\%$ deviations from baseline are detectable, the planet-induced event 
could last as long as $\sim 8$~days.
The peak magnification induced in the source star will be large if the
distance of closest approach between the planet and background star is 
small.

One goal of an observing program would be to discover any lensing
events generated by wide-orbit planets. In addition, by measuring the
magnification as any discovered planet-lens events progress, we can gather enough
data to perform a model fit and to extract the gravitational
mass of the planet. Furthermore, should any of the planets themselves be 
binaries, this could affect the lensing signature in a distinctive way. 

These goals can be achieved in a two-stage process. The first step
is monitoring designed to discover evidence of lensing. The second
step occurs only if it appears that a lensing event is in progress;
in this case, the observing plan can be made more intensive
for a limited time interval in order to track the progress of the event.

\smallskip

\noindent {\bf (1) Monitoring to discover wide planet-lens events:}  
It is important to make this stage as easy as possible,
because there is a long time (in this case years) 
during which there is a low but significant  
probability  
of detecting wide-orbit planets, should they exist.
Fortunately the fact that it takes so long for the lens star to travel
$\sim 20\, \theta_E$ means that it can take a few days for a planet-lens
event to go from start to finish. Monitoring several times a day is 
therefore adequate.
Ideally, there should 
not be gaps in coverage; this favors developing a team 
that can conduct observations across longitudes. If a large
enough number of telescopes are involved, each can take observations
several times a week. With such infrequent observations by individual 
teams, it becomes important that the data
be quickly communicated to one group for each event, in order to identify 
trends as they occur for the purposes of calling
alerts when needed.   

An advantage of a long baseline of observations
  is the opportunity it provides to characterize the intrinsic variability
of both the lens star and the source to be lensed. 
We will discuss this more for the specific case of VB~10 in the next section;
VB~10 provides a clear lesson that it can be advantageous to understand the  
variability in different wavebands. 
   
\smallskip

\noindent {\bf (2) Tracking an event:} Once an event is detected,
it is important to obtain enough data to provide a model fit.
If an event can be identified as such before the magnification has risen by
about $\sim 34\%,$ then more intensive monitoring will be able to 
catch the event during the interval in which it is most variable.
During this interval, a total rate of observations on the
order of every few hours would be adequate for a Jupiter-mass planet. 
An alert can encourage more
observers to join in, providing even more frequent coverage over an interval
of a few days. This could prove helpful if the planet happens to be 
less massive than Jupiter. 

The teams observing predicted events have a number of decisions to make as they design 
a program to monitor a specific predicted event.
For example, what is the minimum mass to which they would like to be sensitive?
Because the time duration scales as the square root of the mass, the
complementary question is: What is an appropriate investment of
resources? To decide the answers to such question it can be
helpful to consider the evolution of the event probability with time.
The  
probability is $\sim1\%$ to detect a planet with $\alpha=20$ (see Table 1), 
at $|t-t_0|\sim500$~days.
As time goes on, lensing by planets with smaller values of $\alpha$ 
becomes detectable, and the probability increases. For example, 
at $|t-t_0|\sim35$~days we there is a probability of $\sim11\%$ of detecting a planet 
with $\alpha=2$.

\begin{table}[h]
\centering
\label{tab}
\begin{tabular}{| c | c | c | c |}
\hline
$\alpha$ & $|t-t_0|$ & $P(A>1.06)$ & $P(A>1.01)$ \\ \hline
2 & $\sim35$ & 0.11 & 0.18 \\ \hline
3 & $\sim80$ & 0.07 & 0.12 \\ \hline
5 & $\sim115$ & 0.03 & 0.05 \\ \hline
10 & $\sim250$ & 0.01 & 0.02 \\ \hline
20 & $\sim500$ & 0.01 & 0.01 \\ \hline
\end{tabular}
\caption{
Probability of an event with
$A>1.06, 1.01$ produced by a Jupiter mass
planetary companion to PMI18537+492,
for a range of orbital 
separations, $\alpha$, as
predicted by Equation (4)
of Paper~I.
This is for a circular face-on orbit; different eccentricities and orientations
will produce different probability distributions.}
\end{table}


\subsubsection{Intermediate times}

In the interval within about $80$~days of the closest passage, 
a greater variety of events can be 
detected\footnote{We define ``intermediate times'' to refer
to the interval which starts when close-orbit planets can be detected. In
principle, this interval can start earlier, if highly sensitive
photometric observations are possible. To be conservative, here we
take the interval to begin about $80$~days prior to closest 
approach, when close binaries with $\alpha=3$ produce a strong signal.}.
Detection of planets in certain types of orbits is all but guaranteed,
should they exist. A well-designed monitoring program will therefore
 either make 
discoveries or place strong limits of the presence of planets with orbits
characterized by $\alpha < 2.$   

By the time the stellar
lens has come within  
about $3\, \theta_E$ of the background source,
four things, summarized below, have changed. 

(1) The probability of events generated by wide-orbit planets has increased.
The 
increase in probability is shown in Table 1 and
 can be seen in Figure 1: as time goes on and
the ruler discussed in the introduction of \S 2.2 moves to
the right, the density of points lining the
 magenta curve increases.   
The increase in
probability is partly due to the smaller size of the orbit (see Equation 4 
of Paper~I) and is partly due to the somewhat larger size of the
isomagnification contours associated with detectable events. 
Furthermore, with a larger event probability,
every day a planet is not discovered
translates into a significant probability that a planet with a given
value of $\alpha$ doesn't exist. 

(2) Events caused by wide-orbit planets are more likely to
exhibit binary effects. This is because the stretching
of the isomagnification contours is accompanied by other distortions,
making the characteristics of wide-planet-lens events more likely to
exhibit deviations from the pure point-lens form. 
One enhanced binary effect is 
that the caustic structures are larger. 
An example of a wide-orbit-planet light curve
exhibiting binary effects is shown in the
red light curve ($\alpha=2.73$)  on the lower-left of Figure 2. 
The deviation of this light curve from the underlying green light curve
is caused by a close approach of the Jupiter-mass planet to the background star.
Although it is difficult to see on this scale, this event actually exhibits
a caustic crossing. 

(3) The 
magnification due to the stellar lens has increased. 
At $\alpha=3,$ the magnification
is about $1.7\%$, potentially detectable. It is convenient to
define $\delta=(A-1).$ From the time the 
value of $\delta > 0$ becomes measurable, the exact 
distance between the lens star
and the background star is known, in units of $\theta_E.$ Each measurement
of the magnification can now be used not only to discover planets, but
also to measure the mass of the lens. 

(4) It has become possible to detect  
close-orbit planets. In the top panel of Figure~2, 
the green deviation in the light curve
at about $390$~days ($\sim 25$ days after closest approach)  
is caused by a close-orbit planet.
A larger set of such deviations is shown in the bottom panel.
More light curves for close-orbit
planets are shown in Paper~I. The key point to 
take away from these images is that the phases chosen to 
generate the light curves were random, yet every light curve
exhibits deviations. For the case of Jupiter-mass planets orbiting
low-mass dwarfs like PMI18537+492 and VB~10, with $\alpha$ approximately
equal to $1/3,$ these deviations can be easily detected. 
Furthermore, the values of the orbital separation and mass ratio can
be determined. Below we summarize results presented in 
\rd\ (2011). 

For each value of $\delta,$ observations are sensitive to planets with
$\alpha=0.84\, \delta^{0.25}$. If such a planet exists, it will produce
a deviation in the light curve like those that shown in Figure 2, for
$log_{10}(A-1)=log_{10}(0.25)= -1.6.$ The relatively short orbital period of
about $26$ days, means that the deviations will be repetitive, although
their size is maximized in the region just mentioned above. 
If $P_{orb}$ the planet's orbital period then 
the time duration of a deviation from the point-lens form is
\begin{equation} 
T_{dev} \approx
2.5\, P_{orb}\,  10^{(0.5\, Log_{10}(q) - 0.2)}   
\end{equation}
This predicts a deviation duration of about $2$~days
for a Jupiter-mass planet in an  orbit with $\alpha=3.$; 
this is reflected in the
light curves of Figure 2. 
  
{\sl Thus, by deciding on a sampling frequency and the depth of the observations
for each measured value of $\delta,$ observers are certain to either detect 
lensing signatures of close-orbit planets, or else to be able to determine
with certainty that certain orbital separations and mass ratios can be
ruled out. 
}
\smallskip

\noindent {\bf (1) Monitoring:}  
Once $\delta$ is larger than about $0.017,$
the monitoring program designed to allow event alerts should be
stepped up. This is because the probability of detecting certain close-orbit
planets, if they exist, approaches unity. Therefore, at this point in
the program it will be productive to increase the numbers of telescopes
in use, and to attempt observations two times per night with each, if possible.
Fast and efficient calls of alerts are needed, because the deviations from the
underlying point-lens light curve associated with lensing by the 
high-proper-motion star, may exhibit more structure than the 
planet-lens events expected
at early and late times.  
   
\smallskip

\noindent {\bf (2) Tracking an event:} The basic principle is the
same as for tracking events that occur at early or late times.
But, because the light curves will tend to exhibit 
structure on short time scales,
more frequent observations after the alert would be fruitful.
This portion of the observing plan would be very similar to plans
carried out to discover planets in present-day microlensing 
programs.

\subsubsection{Times near the time of closest approach}
Once the the stellar
lens passes within  
about $2\, \theta_E$ of the background source (within $\sim 35$~days
of closest approach) we can 
start to see features of planets in the zone for resonant
lensing, in addition to continuing to find evidence
for close-orbit and wide-orbit planets.
During the interval from $35$~days prior to closest approach until
$35$~days after closest approach we are almost certain to discover evidence
of planets in the \zrl , if they exist. Sensitivity to close-orbit planets
continues during this time as well. 

\smallskip{\bf Monitoring and Tracking:} 
The program for monitoring to
discover events and tracking the evolution of any events discovered 
should be the same as described above for close-orbit planets. 
During the time around closest approach, however, it is 
especially important to have frequent, deep monitoring: (a)~close-orbit planets
can cause deviations in the magnification as the lens star wobbles
in its orbit on short time scales (Figure~4 of Paper~I); 
(b)~wide-orbit planets cause 
subtle deviations from the point-lens form in the magnification produced
by the central star, also associated with stellar wobble (Figure~5 of Paper~I).  

\section{Monitoring VB~10}
\subsection{VB~10 as an example of a nearby lens}

VB~10 is 
a nearby high proper motion M dwarf,
whose upcoming predicted close approach with a background star 
was discussed in detail in L\'epine \& \rd\ 2011 and in Paper I.
VB~10 has an estimated mass of $0.075M\odot$, at 
a distance from us of $5.82$~pc, with a proper motion of
about $1.5^{\prime\prime} {\rm yr}^{-1}$. 
These parameters allow us to compute the value of the Einstein angle,
$\theta_E \approx 10$~mas, and the Einstein diameter
crossing time, $\tau_E \approx 5$~days. 

As we did for PMI18537+4929, we conducted simulations considering a
hypothetical Jupiter-mass planet orbiting VB~10  
in a face-on, circular
orbit. 
For VB~10 we selected a physical path compatible with astrometric 
measurements of the system, specifically, the path with 
$\beta=0.5$ ($5$~mas), shown in Figure~1 of Paper I.   
 The resulting pattern of points in the plane of $\alpha-(t-t_0)$
is shown in Figure 3. The most obvious
difference from PMI18537+4929 as shown in Figure 1,
 is the significantly shorter timescale of the event.

The reason the time scale for VB~10's approach is almost ten times shorter than
that of PMI18537+4929, is VB~10's relatively high 
angular  speed. To quantify the comparison with other nearby stars,
we considered $\sim 7500$ stars that happen to
lie 
in the {\sl Kepler} field and which have measured proper motion. 
Setting the value of $D_L/D_S$ to zero, 
we computed the distributions of the values of both $\theta_E$
and $\tau_E.$ Figure 4 shows that both PMI18537+4929 and VB~10 have
large Einstein rings, but that neither is among the largest for nearby stars.
Yet, while the value of $\tau_E$ for PMI18537+4929 is fairly typical of values
for other nearby stars, the value of $\tau_E$ for VB~10 is smaller than
for any star in the sample. This is because it is nearby and also because
its transverse speed of about $40$~km~s$^{-1}$ is also fairly large.   
  
\subsection{Observing the first Predicted Mesolensing Event}

The relative brightness of VB~10 at long wavelengths and its variability
(Cutri et al. 2003; Berger et al. 2008; West et al. 2008; Hilton et al. 2010)
 pose challenges. VB~10 is an active M~dwarf and is known to exhibit flares.
It is important to be able to distinguish lensing signatures from possible flare activity.
For this we must rely on  multi-waveband observations
and on the pattern of time variability. Flares have a charateristic fast rise and
exponential decay. Although rotation effects can alter the light curve (Berger et al. 2008),
it should be rare for a flare to mimic a lensing event. Furthermore,
the time scale of most flares is shorter than that of the lensing events we have considered.
Thus, should we detect evidence of lensing, we expect to be able to distinguish it
from flare activity; this is why we emphasize the need for good time coverage and
mutiband observations. 
In fact, during late November we were able to compare
MEarth I-band light curves with a small number of 7-band observations with 
GROND\footnote{See Greiner et al. (2008) for details of the GROND program.}.
This showed that, even when significant I-band variability of VB~10 was exhibited, the variability was less evident or not evident 
at shorter wavelengths.

The primary challenge we face in designing a program to monitor VB~10
is that the date and distance of closest approach
are uncertain (L\'epine \& Di\thinspace Stefano 2011; 
Paper~I).
{Despite the difficulties, 
we can still learn from the VB~10 lensing event, but what we 
learn
depends on the time of closest approach.}  
{\sl It is important to note that, although we do not know at the time
this text is being written what the distance and time of
closest approach was or will be, future high-resolution images
will allow us to determine the relative path of VB~10 and [VB~10]-PMLS-1
with small uncertainties. 
}

{\bf (1) If the closest
approach has already occurred}, then we may be able to discover lensing
signatures from the time of
closest approach in existing data, such as the combination of MEarth and GROND
data mentioned above. 
Since, however, dense monitoring has not yet begun, and VB~10 is still
near the Sun, 
we are not likely to detect lensing signatures associated with the
closest approach. This means that we will not measure the mass of VB~10,
nor will we be sensitive to signatures from planets in the resonant zone. 
The signature most likely to be detected, however, 
is that due to wide-orbit planets. Monitoring that starts now
and continues for a few months will either detect or place weak limits 
on the presence of wide-orbit planets. 
On every day from now onward, the detection or failure to detect
a planet-lens signature either discovers or places a 
weak limit on the presence 
of a planet with a well defined value of $\alpha.$

In analogy to the work described in \S2.2, we 
can estimate the probability, $P$, that an event with
$A>1.06$ will be produced by a Jupiter-mass planet on a wide 
orbit. 
While simulations like those that produced the
probability plots of Paper I can provide better estimates,
Equation 4 of Paper I provides a reasonable guide to 
the $\alpha$
dependence of of $P$, especially for small $\beta.$\
The probability is $\sim1\%$ to detect a planet with $\alpha=20$ 
at $|t-t_0|\sim46$~days.
As time goes on, lensing by planets with smaller values of $\alpha$ 
becomes detectable, and the probability increases. For example, 
at $|t-t_0|\sim5$~days we have a $\sim27\%$ chance of detecting a planet 
with $\alpha=2$. The increased probabilities compared
to PMI18537+4729 reflect the higher mass ratio for a Jupiter-mass planet orbiting
VB~10.
 
\begin{table}[h]
\centering
\label{tab}
\begin{tabular}{| c | c | c | c |}
\hline
$\alpha$ & $|t-t_0| (days)$ & $P(A>1.06)$ & $P(A>1.01)$ \\ \hline
2 & $\sim5$ & 0.27 &  0.47 \\ \hline
3 & $\sim7$ & 0.16 & 0.28 \\ \hline
5 & $\sim11$ & 0.08 & 0.14 \\ \hline
10 & $\sim23$ & 0.02 & 0.03 \\ \hline
20 & $\sim46$ & 0.01 & 0.01 \\ \hline
\end{tabular}
\caption{
Probability of an event produced
by a Jupiter mass companion to VB~10.
with
$A>1.06,1.01$ for a range of orbital 
separations, $\alpha$, as
predicted by Equation (4)
of Paper~I.
}
\end{table}

{\bf (2) If the approach occurs in the near future}, we are unlikely to
have ideal monitoring during the event. Nevertheless, if VB~10 is orbited by
a planet in the zone for resonant lensing, and if the distance of closest 
approach lies within about $20$~mas, we have a good chance of
detecting evidence of the planet. 
This is because the magnification of the [VB~10]-PMLS-1
will be high enough that we will receive roughly as much light from
it in B band as we receive from VB~10 itself. 
If VB~10 is orbited by a close-orbit
planet, we may be able to detect evidence of it, but only when
observing conditions become more favorable, because close-orbit planets 
will produce deviations that are typically small and, for VB~10, short-lived
as well.  
The main channel of detection will be in the wide regime.
Fortunately, since we will have continuous monitoring from the time
of the event, Table 2 shows that we will be able to place meaningful
limits for modest values of $\alpha$, since the probability that such
planets produce events is high.

{\bf (3) If the event occurs in late winter or early spring}, 
we will be able to
learn a good deal. It is therefore important  
to take high-resolution images as
soon as possible, to refine the prediction, as in this case,
we should enlist a large number of observers. 
{\sl Close-orbit planets:} The orbital period can be very short.
For example, with $\alpha=1/3,$ $P_{orb}=3.6$~days, so that orbital
motion can increase the detection probability to nearly unity.
Equation 2 shows that the deviation duration
would be $\sim 0.65$~days for 
a Jupiter-mass
planet orbiting VB~10 with $\alpha=1/3.$ The fractional deviation due to
the magnification of the dim background star is small, so that
deep observations are required. 
{\sl Planets in the \zrl :} Orbital motion is still important enough to 
significantly increase the probability of detection (see Figure~3 of 
Paper~I). The magnification tends to be large, making it easier to
identify the magnification of the dim background star.   
As is the case for close-orbit planets,
frequent observations will either discover
planets in the \zrl , or else will place 
strong limits on their existence. 
{\sl Wide-orbit planets:} 
There is  a high probability of discovering wide-orbit planets
 during an interval of weeks
from the closest approach.

\noindent{\bf Observing Strategy:} 
While intensive monitoring of a system in which the distance and
time of approach is uncertain cannot be justified, 
a discovery would be important.
We therefore propose a
compromise between the scientific potential
 and the uncertainty associated with the event. First we  consider the
questions of detectability in more detail.  
If telescopes at approximately ten locations around the globe can each
observe the source star $2-5$ times per night, a reasonably densely sampled
light curve can be developed. Because the background star is so dim, 
the signal-to-noise ratios required to reliably measure the
magnification may require meter-class telescopes. 
The exception        
would be for high-magnification. 

\smallskip

\noindent{\bf Analysis:} For each value of $\alpha$, a set of orbits are possible. This set
includes face-on orbits in either direction, inclined orbits, and
eccentric orbits. For each type of orbit, and for a wide range of
possible planetary masses, we compute the light curves, specifically
computing the value of the magnification at each time an observation occurs.
This will identify the cases in which we would have had a high or low probability
of planet detection, and may even lead to the discovery of a planet.

Due to the proximity in time of the VB~10 event, we have not yet
conducted a full analysis of orbits in the general case- i.e. including 
eccentricity and inclination effects. Throughout this paper and Part 1 we have considered
circular face on orbits, and will conduct a full general treatment
shortly.

\section{Will event prediction be common?}

Feibelman's identification of the possible 40-Eridani event was
almost certainly serendipitous, as was our discovery that VB~10 would
make such a close approach to a background star. 
The relative ease of these predictions
suggests that lensing event prediction should not be too difficult. 
It is therefore important
to make a quantitative estimate. 

This was 
first done by Paczy\'nski (1995), who calculated the area of the region
swept out per year
within $\theta_E$ of a typical nearby lens within $D_L=10$~pc.
He found that, if the background source density is $0.3$ per sq. arcsecond, 
then each
nearby dwarf should produce, on average, slightly
fewer than one event per century. Thus, by monitoring a few hundred
of the nearest stars, one can expect that a few of them per year will 
produce events in which the background star is magnified by
$\sim 34\%$ or more.  
Paczy\'nski (1996) later went on to consider events displaying
the astrometric 
effects of lensing, for which the distance of closest approach can
be larger, leading to a higher event rate. 

Here we include for the first time, lensing by possible planetary
companions to nearby stars.  
Every time a nearby star produces an event, there is an opportunity to
discover planets in the zone for resonant lensing as
well as close-orbit planets. In addition, the stellar-lens light curve
may be distorted by the fact that the lens star is wobbling in response to
an orbiting wide-orbit planet.  
In addition to possibly being discovered through their influence on
the motion of the star they orbit, when it serves as a lens, 
wide orbit planets can  
 be discovered by producing independent events that may or may not
be preceded or followed by a detectable stellar-lens event.
If a 
population of $N$ nearby stars is being monitored, and each
produces events at a rate ${\cal R}_j,$ where $j$ labels the star, then 
the rate of independent wide-planet-lens events is  
\begin{equation} 
{\cal R}_{wide}^{ind}=\sum_{j=1}^N\, {\cal R}_j\,  
\sum_{i=1}^{n(j)} \sqrt{\frac{m_{p,i}}{M_{*,j}}}, 
\end{equation}
where $i$ labels the planets associated
with an individual star, and   
$n(j)$ is the number of planets orbiting the $j\,th$ star. 
The rate of separate planet-lens events can be a significant
fraction, as much as several tenths, of the rate at which
nearby  stars produce events, depending on the number of planets per
star and the mass distribution of the planets.  

We also extend the reach of these studies by considering lensing by somewhat more
distant stars. While the proper motion of stars more than $\sim 10$~pc
away is generally
smaller, producing a smaller event rate per star, there are
nevertheless a large number of stars within a hundred parsecs or so that
have have proper motions larger than $0.15$~arcseconds per year.

First we consider the rate at which nearby stars produce events,
then turn to the enterprise of prediction. 
To determine 
the event rate, we have considered
approximately $7500$ stars with measured proper motion
that happen to lie in the field observed by the {\sl Kepler}
space mission. For each of these stars we have estimates of
the mass and distance, in addition to the proper motion. This 
allows us to compute the size of the Einstein ring of each star and the
area covered by the Einstein ring per year.   
We find that, in total,
 the Einstein rings of these stars cover approximately 
$3.6$~arcseconds per year. 
Events can be detectable even when the angle of closest approach is
larger than $\theta_E.$ For example, deviations 
of  
$6\%$ or more are likely to be detectable, so that the 
angle of closest approach need only
be smaller than 
$2\, \theta_E.$ Thus, the lensing region of nearby stars in the {\sl Kepler}
field covers about $7.2$ sq.\, arcseconds per year. Over the whole sky
(which has about $400$ times the area of the {\sl Kepler} 
field) the lensing regions of nearby stars cover approximately 
$3000$~sq.\, arcseconds per year. If the background stellar density is 
$0.3$ per sq.\, arcsecond, then we expect roughly $1000$ events across the
sky caused by nearby lenses.

These events will be generated by over $2\times 10^6$ stars,
most of which have already been identified. The question then
becomes, how do we make predictions? 
There are two types of answers to this. The first is to simply develop an
ensemble of possible lenses and to arrange to monitor enough of them
that we are guaranteed to detect several events per year. We will  
know which specific monitored stars produce the events only after 
the events begin. 
Wide-field surveys that are more-or-less automated can simply keep track
of any changes in the amount of light received from the positions around
all high-proper-motion (HPM) stars. 

Surveys that must select a limited number
of directions to monitor must know which of the high-proper-motion
stars are most likely to produce events.
This can be discovered by computing, for each HPM star, the  
area it covers in the sky. Those that cover the largest area are
generally the best candidates to produce events.
By studying the sample of stars in the 
{\sl Kepler} field, we find that a small fraction
of the HPM stars have a very high probability of producing
lensing events:
$0.6\%$ of all of the possible lenses will likely produce 
$10\%$ of the events. The most likely lenses tend
to lie within about $100$~pc. Thus, subsets  of high-probability lenses
can easily be selected; in fact we have already done this for the {\sl Kepler}
field (\rd\ et al. 2012b).

These numbers  make it clear that it is possible to identify 
ensembles of nearby stars, each of which has a high probability
of producing a lensing event sometime during the next year. In fact,
the total probability is so large that, by monitoring these stars,
we are guaranteed to discover dozens of event per year.
The reason these events are special is that each can yield the 
gravitational mass of the lens star and can explore its planetary
system in detail, potentially discovering several planets per star
and measuring the mass and orbital separation of each. 
It is clear that, if we focus attention on predicted events, 
we will have a tool that can be continuously employed to study nearby 
planetary systems.  

It is not necessary to rely on only statistical probabilities. 
We can identify potential lens-source pairs
by using catalogs 
of nearby stars
with measured proper motion, and by cross correlating them with catalogs
of nearby background stars.
In fact catalog matches were suggested by
Feibelman\footnote{Feibelman wrote (1986) ``It is hoped that this 20-year
exercise in frustration will encourage others to conduct
systematic searches, perhaps by means of computer-based data banks,
for stars with large proper motions that eventually may eclipse a
background star and give rise to the elusive gravitational lens effect.''},  
and implemented by Salim \& Gould (2000), who
 used catalogs to identify pairs of nearby stars
that could induce astrometric shifts in more distant stars. The idea
was to use the then-proposed {\sl Space Interferometry Mission} (SIM)
to observe these events,
due to its excellent astrometric precision of $\sim 4 \mu$~arcsec. 

This discussion leads us to develop a hierarchy of prediction.
{\sl Level 1:} At the lowest level are the stars with measured proper motion, 
such as those in the LSPM catalog (L\'epine 2005; L\'epine \& Shara 2005; L\'epine et al. 2002, 2003).
Large-scale semi-automated 
surveys like PTF and Pan-STARRS  can keep track of the light curves
of each such star in their fields. 
{\sl Level 2:} In order to select the best stars to monitor for smaller
or less automated surveys, or to decide on a set of fields that might
receive highest priority in large surveys, it is important to know which stars
have the highest probability of producing lensing events.  
To identify them, we must be able to estimate the
masses of the stars with measured proper motion, and the distances to them.
Using this information, we can identify those 
whose Einstein rings cut
the largest swath across the sky. If, for example, it is
feasible to monitor $2000$ stars, we may decide to select the
top $2000$ stars from this list.  
{\sl Level 3:} At level 3, we consider the background stellar density.
Stars that cut a smaller swath across the sky may nevertheless have a higher
probability of producing events if the density of stars behind them
is larger. Thus, level 3 is simply a reordering of the 
systems in level 2, based on a more realistic estimate of the 
event rate per star. {\sl Level 4:} We consider all of the stars in level 1
and identify those for which there are high-resolution images of  
the background extending over time. For this set of 
systems, we can compute the path of the potential lens and identify
stars that lie close to it. In cases in which a full astrometric
solution is possible, we can predict the approximate distance and time
of closest approach. If there are too many images to analyze, we can 
choose to start with the potential lenses at the top of the level 3 list, since  
they have the highest probabilities of creating
lensing events.  

At the end of this  process we should be able to predict a 
set of specific events.   
Because the number of such events per year depends not just on the
properties of nearby stars and the stellar fields in front of which they travel,
but also on the availability of data, it is not possible to predict how many
individual events can be predicted each year. It seems likely that
at least a handful of individual photometric events will be predicted each year.
The ability to detect astrometric events with a mission such as Gaia,
should lead to a larger rate of successful event predictions. 
Since the elements of day-by-day monitoring for individual events
were covered in sections 2 and 3,
we turn below to the issue of monitoring ensembles of predicted events.

\section{Monitoring Ensembles of Sites of Predicted Events}

Once an ensemble of potential future lensing events is identified, 
monitoring to detect evidence of lensing can begin.
The ideal approach is
a partnership among several groups. This 
minimizes gaps in coverage, while reducing the telescope time
needed from any single group.  
It is important to 
have a central clearinghouse that receives data soon after each 
observation, so that photometric changes can be detected
quickly enough to call an alert when lensing events start.

At present, alerts are called by lensing monitoring teams so that almost continuous 
coverage of a lensed star can be achieved during a period when the magnification
is changing rapidly due to planet-lens effects (see e.g, Janczak et al. 2010; Miyake et al. 2011).
The need for continuous observations will
be rare for nearby lenses, since effects caused by nearby planets tend to
be longer-lived because of the relatively large sizes of the 
Einstein rings (Figure 4).
Instead, an alert will be designed to encourage more observers to join in
to achieve monitoring as frequent as a few times per hour (i.e., a few times per
day per team) over an interval of several days. 
With this type of observational plan in mind, the discussion below
explores how different observing teams can begin the search for 
lensing by an ensemble of stars, each 
having a high probability of causing lensing events in the near future. 

\subsection{Wide-Field Surveys} 

With $100-1000$ nearby-lens events per year, and a detection
efficiency likely to be smaller than unity, it is important to
survey large regions. Fortunately,   
wide-field surveys are already being conducted.
Some of the teams conducting wide-field monitoring are large professional
programs. These include PTF and Pan-STARRS. Others are programs of
long-standing run by amateurs.\footnote{For example, an amateur program
that surveys $4500$~sq. degrees every two nights to find novae was the
first to discover a mesolensing event in the field. The Tago event, named
for its discoverer, involved a nearby lens that passed in front of an A0 star
1~kpc away (Fukui et al. 2007; Gaudi et al. 2008a).} 
These surveys differ from each other in
the photometric sensitivities they achieve and in the 
observing cadences they employ. None of the
wide-field surveys currently operating are searching for lensing across the 
sky\footnote{Pan-STARRS has a sub-program {\sl Pandromeda} that has
identified lensing candidates in M31.}. The quarry considered most exciting
by these teams are explosive transients. 
Lensing is 
one of many other types of less dramatic variability,
and the existing teams have limited resources
for research outside their main purview.     

Event prediction will allow the teams
to sidestep the difficult task 
of characterizing variability
and identifying lensing events among the much larger set of
all variables.
Simply by supplying these teams with the coordinates 
of the stars likely to cause events within the next few years, we can
make it possible for them to begin identifying event candidates
immediately. 
Such a list is now relatively
straightforward to develop. Indeed, our group has already identified 
several hundred high-priority 
sites to monitor for the {\sl Kepler} space mission (\S 5.5; \rd\, et al. 2012b).
.
We also plan to publish an all-sky table of the sites of predicted events,
prioritized according to levels 2 and 3.

The light curves from these locations should be carefully checked
with each observation 
to assess the
probability that variability at 
the site of a predicted event is due to lensing.
Because observations by the wide-field monitoring surveys are 
taken regularly during an interval of years, they have the 
advantage of developing a long baseline. Information collected over the
this timescale can help to determine if
variability at the site of a predicted event 
is unusual in terms of its magnitude,
color, or temporal pattern.

One caveat is that most fields covered by these surveys are 
observed at intervals of several days, which is too
infrequent. Since, however,  several surveys  
may be observing the same sites of predicted events, a clearinghouse can
compile a composite light curve that can be used to identify
lensing-like signatures on a shorter time scale. 
Once a system is determined to  be possibly experiencing lensing,
additional observations are needed. These can test the lensing hypothesis
and measure the mass of the lens star and/or of its planets. 
We address the issue of follow-up in \S 5.4.

\subsection{Ongoing Lensing Monitoring Programs}   

Ongoing lensing teams are of course the best equipped surveys to discover
and identify lensing events. At present, however, it is difficult to predict
specific events for two reasons, both related to the fact that these
programs monitor dense stellar fields. First, it is difficult to
measure the proper motion of stars traveling across regions with a high
surface density of stars, and values found
in catalogs are often not reliable\footnote{In a preliminary study that checked
images against the published proper motion of stars near lensing events, we
found that the images could fail to confirm the result, presumably
because variability in a dense stellar field can mimic stellar motion
(McCandlish et al. 2010; McCandlish et al. 2012).}. 
This problem can be overcome by  
the monitoring teams themselves: with a long baseline, they can
measure the proper motion of nearby stars against the background of distant
stars. This has so far only been carried out for part of the data 
(Alcock. et al. 2011; Rattenbury et al. 2008).
Second, the background stars are highly blended, making it difficult to    
know which specific star might serve as a lensed source.

Nevertheless, we know that some of the lensing events discovered by the
monitoring programs are generated by nearby lenses. At present,   
lensing teams like OGLE and MOA are monitoring large regions of the 
sky, on the order of 
$200$~sq. degrees, or $\sim 0.005$ of the sky.
If, therefore, there are $1000$ events per year
generated by stars within about $100$~pc, roughly $5$
per year
of them occur in the fields monitored for evidence of lensing.
Although these events may not have been predicted in advance, they are
ideal for the types of study outlined for individual
predicted events in \S 2 and \S 3, because we know that $\beta$ is small
and that the background star is bright enough to render the rest of the
event detectable. 
In particular, the proximity of the lens means that  
$R_E$ is likely to be small, and planets
in orbits with small values of $\alpha$, even with $\alpha$ near unity,
will have short orbital periods. This means that, if we conduct frequent
enough monitoring, and if we can detect 
small changes in the amount of light received
from the lensed source, we will either discover planets or else will know
that there are no planets within about $2\, \theta_E$ of the lens star. 
There is also a significant probability of discovering wide-orbit planets, 
if they exist, with a day-by-day monitoring program either finding 
wide-orbit planets, or
placing limits on the existence of planets in well-defined orbits.  
In addition, the value of $\theta_E$ will be large enough to minimize or
eliminate finite-source-size effects. 

If therefore an event  
posted online in real time by the 
monitoring teams\footnote{http://ogle.astrouw.edu.pl/; http://www.phys.canterbury.ac.nz/moa/}
can be identified as being due to a nearby lens,
we can immediately start additional observations which can test for
close-orbit planets and planets in the \zrl . We can also ensure that
these observations continue to supplement those of the lensing teams
after the event, to search for evidence of wide-orbit planets.  
(Please see \S 5.4.) It is important to note that some of the light-curve
features associated with close-orbit planets and/or with planets in the
\zrl , may make the light curve seem different from a typical lensing
light curve. This is because the effects of rotation are to distort
the features familiar from the more common static case. It is important
that the lensing monitoring teams identify these possibly-unusual-looking
events as good lensing candidates, so that they will be targeted 
for further study.   

In addition 
to the handful of events per year generated in the lensing fields by 
stars within about $100$~pc, a much larger number of events are
generated by stars within about a kpc (\rd\ 2008a, 2008b).
In fact, as many as $10\%$ of the $\sim 1500$~ events they post per year,
may be caused by stars within a kpc. While these lenses will have somewhat 
larger values of  $R_E,$ close-orbit planets and wide-orbit 
planets can be found in the manner outlined in \S 2 and \S 3.
The search for planets in the resonant zone can be conducted with 
existing protocols. This has been demonstrated by the discovery, through 
a lensing event that exhibited evidence of orbital motion, of a system
with 2 planets in the \zrl, 
only $1$~kpc away (Gaudi et al. 2008b).  

The key issue is selecting, from among the lens candidates identified by
the monitoring teams, those specific events which happen to
have been caused by nearby lenses. This can be accomplished
using searches through existing catalogs.   
In fact our group has already studied catalog matches to
all of the events discovered that, over a $15$~year interval 
roughly $\sim 8\%$ of the lensing event candidates
have matches to catalogued stars
that could correspond to nearby lenses (McCandlish et al. 2010; McCandlish et al. 2012). 
We are refining the assessment
procedure. Ideally, data from the monitoring teams can also be
used to
determine if there is evidence of a high proper motion star
along the line of sight to each event.

\subsection{MEarth $+$}

\subsubsection{The MEarth Concept}
As described in \S1.2, MEarth monitors 2000 stars across the sky (Irwin et al. 2009a). 
This all-sky coverage is ideal for a survey of predicted lensing events
since the sites of predicted events are scattered across the sky.
MEarth consists of a bank of small telescopes which switch from field
to field during the course of a night. Because the goal is to
discover planets transiting M~dwarfs, MEarth monitors in I band.
If a MEarth target star happens to lens a background star, 
then MEarth could discover lensing events
in the data it  collects. This is most likely to happen if
the background star also happens to be luminous in I band. 
As we see with the VB~10 event, however, the lensed star could provide only a tiny
fraction of the light received in I band, rendering 
a lensing event essentially invisible.
 
\subsubsection{Additions to MEarth}

While the basic design of MEarth is well suited to the study of
the sites of thousands of predicted lensing events spread across the sky,
some alterations would make a MEarth-like project better suited to monitoring
lensing events. If MEarth, or a follow-up project of similar design were to
make the discovery of lensing events by M~dwarfs one of its scientific goals,
it could take several steps to increase the discovery rate.

First, the set of stars to monitor could be chosen with the idea of
maximizing the area of the sky covered by the Einstein rings of the targets.
In fact, the stars already selected are high-proper-motion stars
like VB~10, well suited to lensing studies. As additional stars are
selected for study, the likelihood that each will produce detectable lensing
in the near future could provide an additional selection criterion. 
Useful information for this purpose would include not only the size and speed 
of the dwarf star's Einstein ring, but also information about the   
background stellar field. Higher priority could be assigned to
potential targets whose travels are likely to bring them in front
of stars that can be detectably lensed.  

Second, because the background star to be lensed may not be bright in I band,
multi-waveband observations are necessary. These could be taken by a new
MEarth-like project, if two or more filters can be utilized. They could
also be taken by independent telescope participating in the program
with the idea of supplementing the basic MEarth coverage to improve
the detection efficiency for lensing events. 

Finally, the analysis of the data should include fits for lensing models.
If software suitable to this purpose is applied to existing MEarth data, 
it may find evidence of past lensing events. 
If a typical MEarth target has $\theta_E=10$~mas, and 
$\mu=0.15^{\prime\prime}$~yr$^{-1}$, then over a 5-year period, the
total region producing magnifications of $6\%$ covers a significant
portion of the sky, about $60^{\prime\prime}$. Although not every star is
observed at all times, the area covered by the targets during 
MEarth observations is almost certainly
large enough that some events occurred.

\subsection{Collaborative Observation: Before, During, and After}

The idea of ``follow-up'' is that no single observing program 
can collect all of the data needed
to discover and study planets around nearby stars. It is therefore
important to build a team that works together and
that can also call occasional alerts as needed 
to enlist even more observers.   

Follow-up has been an important part of lensing searches for planets.
During certain events deemed as good candidates for planet searches,
alerts are called to start intensive world-wide monitoring.
Observations that are almost continuous during an interval that is
typically longer than several hours and shorter than a few days 
are taken by 
different groups, using telescopes across longitudes.
The data are combined to provide exquisitely detailed model fits. 

For predicted lensing events, the concept of follow-up must be
generalized. When a specific event, like the VB~10 event, is anticipated,
monitoring must begin before lensing is expected, and continue 
after the time of closest approach. This is needed to discover planets
with a range of orbital periods. Monitoring need not be very frequent:
a few times per night for each of several teams spread across longitudes 
will be sufficient. Once evidence of lensing is detected, 
more frequent coverage may be called
for, depending on the situation. 
In most cases, a total of a few observations per hour, 
conducted by a combination of teams, will be able to resolve the lensing
light curves enough to determine if there are planets and even to discover
multiple planets. In some cases, evidence that a short-lived deviation is 
occurring may trigger intensive monitoring during a limited time interval, similar
to what is required in other planet-lens searches.  
Monitoring in different wavebands is more
likely to avoid the ``dazzling'' effect of the lens.  
When an ensemble of high-probability events is being monitored, the same 
considerations apply. 

Networks 
of observers have already been set up and have successfully
studies a number of different individual systems, including
planetary lenses. These projects have engaged a broad range of
participants, including
the Las Cumbres Observatory Global Telescope Network 
(LCOGT\footnote{http://lcogt.net/}),
the AAVSO\footnote{http://www.aavso.org/} 
and others.
A network for predicted events is presently being 
formed\footnote{https://www.cfa.harvard.edu/$\sim$jmatthews/vb10.html}. 
  
\subsection{Kepler}

The {\sl Kepler} mission monitors approximately $150,000$ stars, with the 
goal of identifying transits by Earth-sized planets.  It was not designed to
identify lensing events, but we have conducted calculations showing that
in the {\sl Kepler} field there are a large number of possible lensing candidates
(see Figure 4). The excellent photometry from {\sl Kepler} means that
small perturbations can be reliably detected. 
Such perturbations could be associated with distant approaches by the lens star,
possibly including deviations by planets in very close orbits.

By conducting a study of the high-proper-motion stars in the {\sl Kepler}
field, we have identified a set of about $700$ stars 
that have the highest probabilities of producing lensing events.
We have created a program to make sure that these stars are included
among the {\sl Kepler} targets. It is almost certain that one of these stars, or 
even some of the other stars already targeted, will produce low-magnification events 
during the expected $3.5$-year duration of the {\sl Kepler} mission (Di\thinspace Stefano et al. 2012b) 

\subsection{Transits and Lensing: A symbiotic connection}

The observations needed to monitor predicted lensing events are well suited
to discovering planetary transits. In fact transit-search programs 
such as MEarth and {\sl Kepler} are sensitive to lensing events by the
target star and/or its planets (\rd\ et al. 2012b), while
the lensing monitoring teams find evidence of transits (Dreizler et al. 2003; Konacki et al. 2005).	
The observations we propose will test each nearby potential lens
for both effects.

The probability that the orbital inclination is favorable for transits
is $R_*/a.$. For a $0.25M_{\odot}$ star with a planet in an orbit of $0.13$~AU, 
corresponding roughly to the center of the habitable zone (Di\thinspace Stefano \& Night 2008), 
this probability is roughly equal to $0.009$ . The number, $N_L,$ of lensing events caused by nearby
a high-proper motion star is 
\begin{equation}
N_L=0.0018\, {\rm yr}^{-1} \bigg(\frac{\theta_E}{10~{\rm mas}}\bigg) 
\bigg(\frac{\mu}{0.15 ~{\rm arcsec~yr}^{-1}}\bigg) \bigg(\frac{\sigma}{0.3~ {\rm arcsec}^{-2}}\bigg)  
\end{equation}
where $\sigma$ is the density of background stars on the sky.
Depending on the time scale of the observations,
the probability of detecting lensing can be comparable to the probability associated with
transits described above.
Every time we conduct monitoring for a predicted lensing event, 
there is an opportunity to look for planets through both their possible
lensing and transit signatures.

\section{PLAN-IT: Prospects}

The idea of PLAN-IT is to plan observations of nearby stars likely to
serve as lenses. This strategy is different from the now common approach
of monitoring many stars in a dense field in hope that an unknown mass
will happen to lens a more distant star. By focusing on nearby stars
that are potential lenses, we are likely to know the proper motion and distance to the lens, so that the light curve fit, which provides an
estimate of the Einstein-diameter crossing time,
 immediately provides an estimate of the
gravitational mass of the nearby star. In addition, we have shown that
planets orbiting the nearby star can produce lensing signatures,
whether or not lensing by the star is detected. When, however, lensing
by the star is detected, tests for planets in all orbital ranges can be
carried out. It is even possible that the astrometric effects of lensing
will be detected in some cases.

The nearby dwarf star VB~10 is the first test case. 
This text is being completed in mid-February 2012. 
If the closest 
approach between VB~10 and the background star [VB~10]-PMLS-1
 has not yet
occurred, and if the approach between VB~10 and the background star [VB~10]-PMLS-1 
is within about two $\theta_E,$ then the upcoming event will 
measure the gravitational mass of VB~10, determine with 
certainty whether it has planets orbiting within about two $\theta_E,$
and quantify the probability that VB~10 has planets in wider orbits.
Even if the closest approach has already occurred, and even if the closest
approach is (or was) more distant, we can still either discover
wide-orbit planets or derive constraints on their presence.  
These constraints will be very weak, but they nevertheless represent
the exercise of a new capability.

Enhanced monitoring of the
region around VB~10 should begin immediately. 
To balance the excitement of a potential discovery with the
uncertainty that detectable lensing will occur, we propose
a modest program: observations from $10$ or more locations
across longitudes, about twice each night from each location. The observations
should be combined as they are taken to identify trends so that 
an alert can be called if 
there is evidence of deviations which may be short-lived. 
Frequent I-band observations of VB~10 with MEarth will soon 
resume, as will
observations with GROND. GROND, a 2-m telescope, provides 7-band coverage
that has a good chance to detect the lensing of the dim blue background
star, [VB~10]-PMLS-1. When new telescopes join in, observations in wavebands
blueward of I will be useful.       

The significance of the observing program extends beyond the VB~10 event. 
Whether or not a discovery is 
made,  
the VB~10 event 
calls our attention to the fact that prediction programs can now be started, and
provides a test case about which to organize the first observing campaign.
A modest investment of a few observations per telescope per 
night, spread out across
longitudes will allow us to not only check for lensing events but
also to test an observing network that can be
called into action for future events.

We expect future events worthy of monitoring to be identified regularly,
by a process that proceeds systematically by starting with catalogs of
high-proper-motion stars. Groups of such stars with the highest
probability of producing events can be identified and monitored by
programs that observe large regions of the sky. These
programs will range from those run by amateur astronomers who
hunt for novae, to large semi-automated programs like PTF and Pan-STARRS,
to programs like MEarth that systematically target interesting systems
spread across the sky.

The stars with the highest probability of producing events in the near
future can be identified through studies of their proper motion that
incorporate cross-correlation with catalogs of possible background stars
and analysis of any existing high-resolution images.    
This process will lead to the identification of individual events
that can be monitored to learn about both the high proper motion star
and any planetary system it harbors (\S 2 and \S 3).   

It is worth noting that the 
sets of observations designed to search for lensing by nearby stars
are also ideally suited to the search for transits by planets
that may orbit these stars in edge-on orientations. Thus, lensing searches
and transit searches are automatically carried out for each star studied.
In fact, the ongoing
 MEarth program could find lensing using its current
strategy. This is most likely to happen if the lensed background star
happens to have a color similar to that of the M dwarfs that they are
targeting for a transit search.

Beyond the serendipitous circumstance of being able to search for planets
in two ways at once, lensing searches for planets around nearby stars are 
intimately connected to other planet-search techniques. This is simply
because these lensing studies will discover planets orbiting nearby
stars or will place quantifiable limits on the masses and orbital separations.
These studies can therefore complement and inform radial velocity and direct
imaging studies. 

In summary, we have proposed a monitoring strategy for VB~10. 
Analogous considerations for other individual predicted events
will shape future programs to optimize what we learn about the 'planetary systems of nearby stars predicted to produce lensing events.
In addition to the monitoring of individual systems, ensembles of
future events can be monitored by a variety of surveys. This type of program
will produce guaranteed results in the form of  discoveries of
planets of a variety 
of masses and with a range of orbital separations     
orbiting specific nearby stars that can be targeted for complementary
observations. 

\bigskip
\noindent {\bf Acknowledgments}
\smallskip

\noindent We would like to thank Christopher Stubbs; Christopher Crockett; Fred Walters;  
David Charbonneau, Zachory Berta and the MEarth project; 
Jochen Greiner at GROND; Matthew Templeton and the AAVSO for their help 
and advice
with observing VB~10. This work was supported in part by NSF under 
AST-0908878. 

\begin{figure}
\centering
\includegraphics[width=1.0\textwidth]
{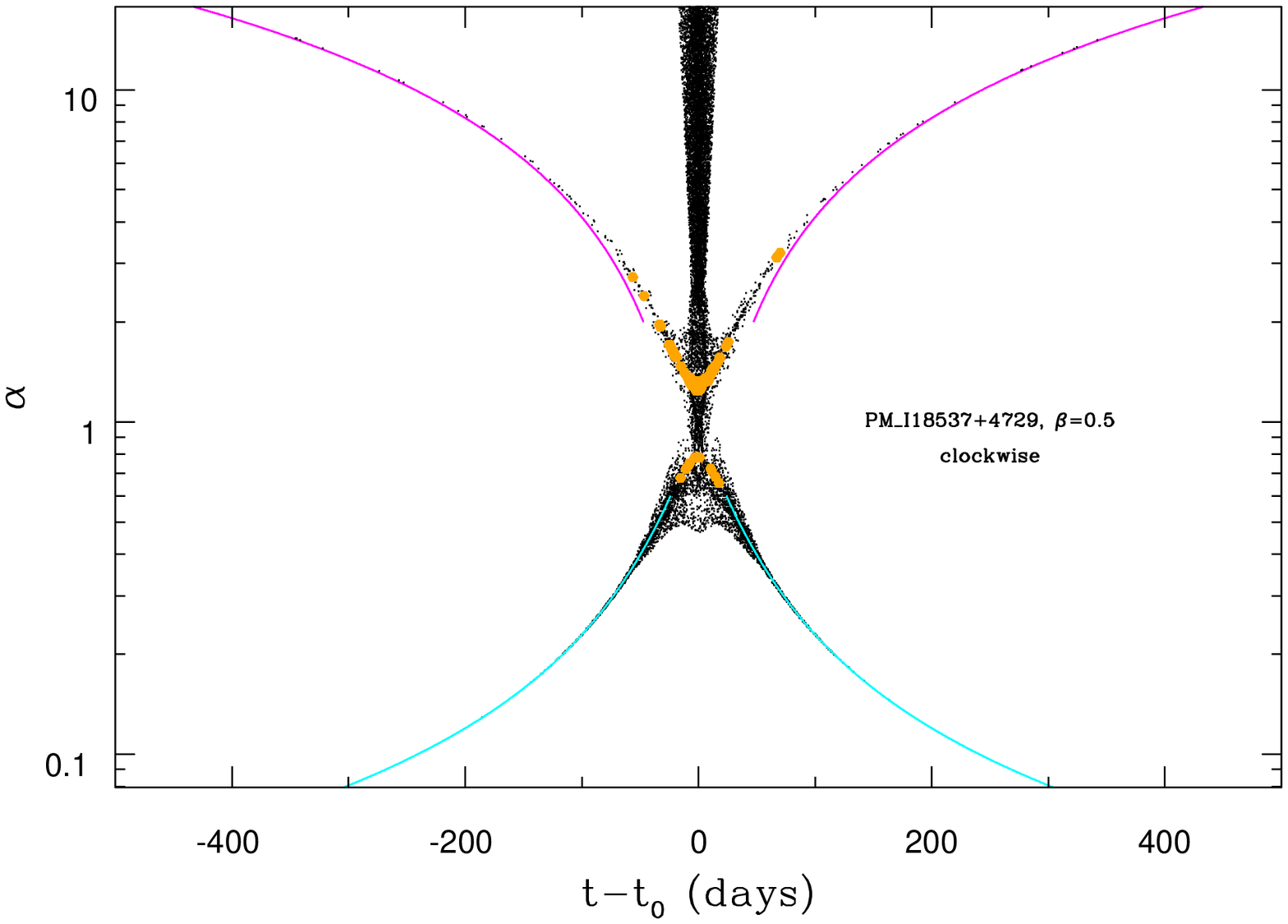}
\caption{
The logarithm of $\alpha,$
the orbital separation in units of $R_E,$  versus $(t-t_0)$, where 
$t_0$ is the time of closest approach. For each point shown, 
$t$  is the time of a peak in the photometric deviation $|A_{pl}-A_{pt}|$ from the single lens;
we consider only those peaks with $|A_{pl}-A_{pt}|>0.01$.
The lens is PMI18537+4729 (see \S2),
orbited by a hypothetical Jupiter-mass planet in a circular face-on orbit.
The distance of closest approach is $\beta=0.5$.
The smooth magenta curves that start in the upper corners of the plot
show the analytic prediction for events produced by wide-orbit planets.
Similarly, the smooth cyan curves that start in the lower corners 
show the analytic prediction for events produced by close-orbit planets.
The orange 
points show the times of events associated with caustic crossings. 
} 
\end{figure}

\begin{figure}
\centering
\includegraphics[width=0.8\textwidth]
{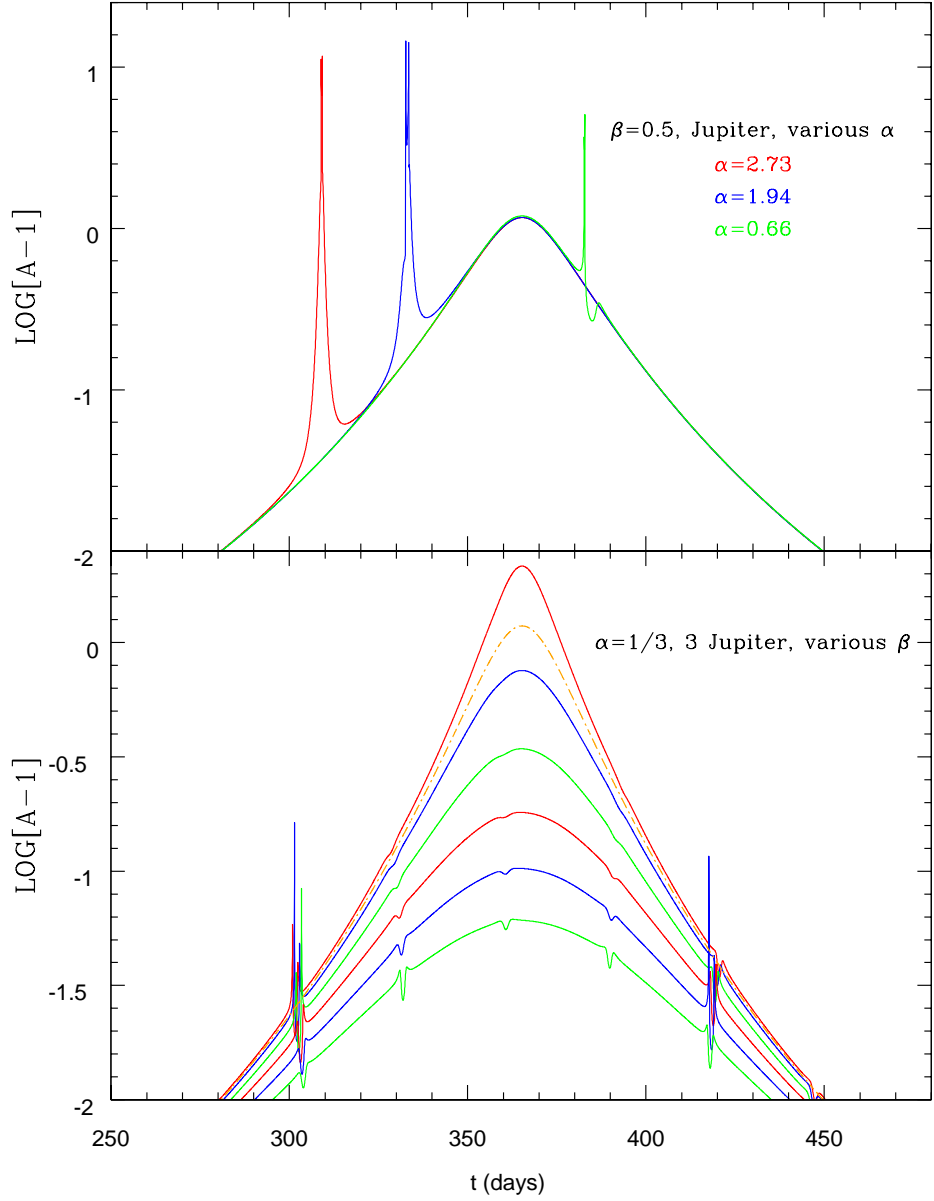}
\caption{
Light curves for hypothetical planets orbiting PMI18537+4729.
{\sl Top Panel:} Light curves exhibiting caustic crossings 
for three different values of 
$\alpha$; in each case, $\beta=0.5$.
{\sl Bottom Panel:} Light curves for $\alpha=1/3$ for a variety 
of different distances of closest approach between PMI18537+4729
and a background star. In this case the planet's mass
is $3\, M_J.$ The orange dashed
curve 
corresponds to $\beta=0.5$, as in Figure 1. The other curves
form a sequence of increasing distance of closest approach, 
with the highest peak having $\beta=1/3$ and $\beta$ increasing
in increments of $1/3$ for each subsequent light curve with 
lower peak magnification.
} 
\end{figure}

\begin{figure}
\label{jup10}
\centering
\includegraphics[width=1.0\textwidth]
{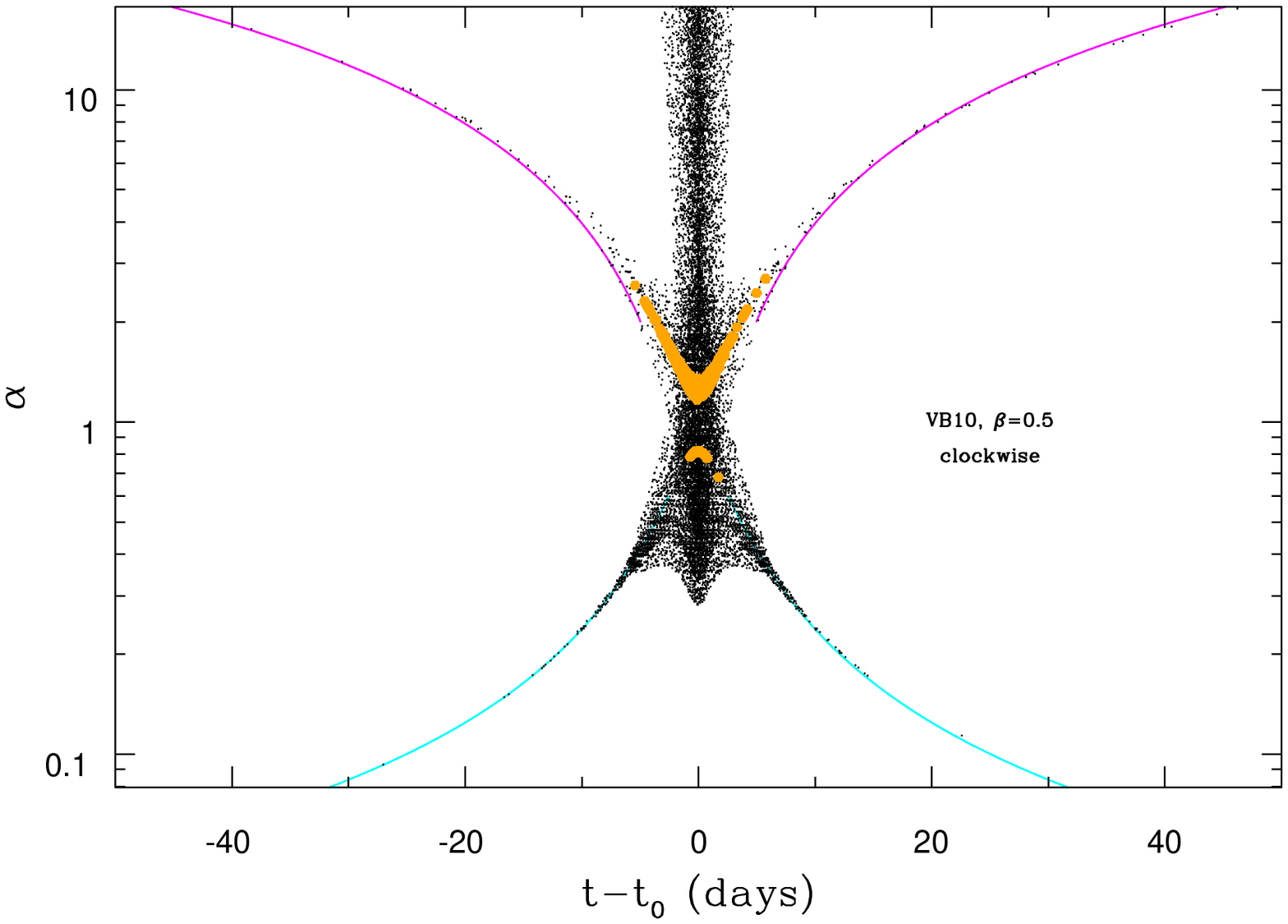}
\caption{
The logarithm of $\alpha,$
the orbital separation in units of $R_E,$  versus $(t-t_0)$, where 
$t_0$ is the time of closest approach. For each point shown, 
$t$  is the time of a peak in the photometric deviation $|A_{pl}-A_{pt}|$ from the single lens;
we consider only those peaks with $|A_{pl}-A_{pt}|>0.01$.
The lens is VB~10 (see \S3),
orbited by a hypothetical Jupiter-mass planet in a circular face-on orbit.
The distance of closest approach is $\beta=0.5$.
The smooth magenta curves that start in the upper corners of the plot
show the analytic prediction for events produced by wide-orbit planets.
Similarly, the smooth cyan curves that start in the lower corners 
show the analytic prediction for events produced by close-orbit planets.
The orange 
points show the times of events associated with caustic crossings. 
} 
\end{figure}

\begin{figure}
\centering
\includegraphics[width=0.8\textwidth]
{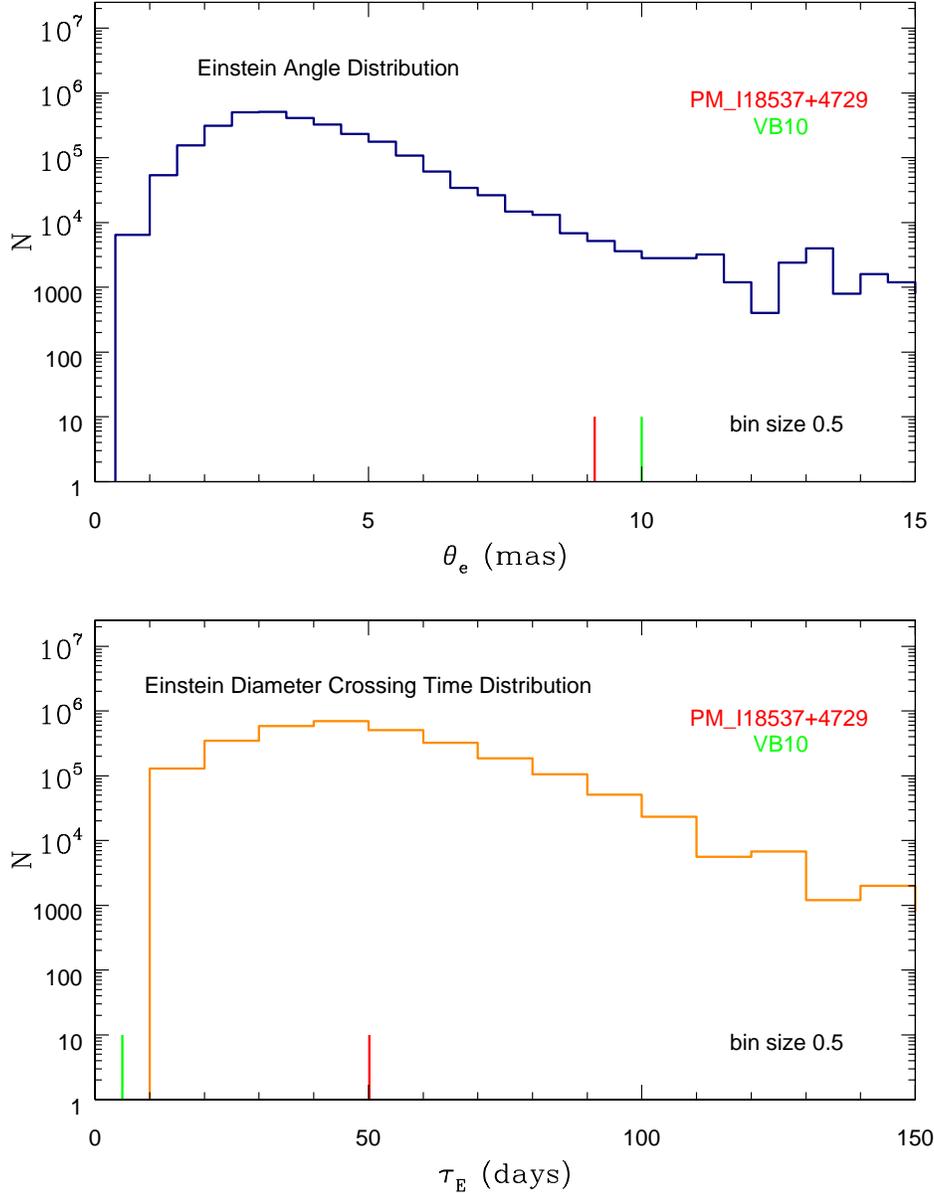}
\caption{
{\sl Top panel:} A histogram showing the distribution of Einstein angles 
across the whole sky
of stars with proper motions $>0.04~{\rm "yr}^{-1}$,
extrapolated from
a sample of 7474 stars across $100$ square degrees in the 
{\sl Kepler} field of view. The values for VB~10 and PMI18537+4729
are marked with green and red lines respectively; the bin
size is $0.5$~mas. These nearby stars are mesolenses, having Einstein rings significantly
larger than typical microlenses, which have values of $\theta_E$ 
often less than a mas. They have high probabilities 
of serving as lenses and may produce detectable astrometric shifts in the
positions of the stars they lens.
{\sl Bottom Panel:} A histogram showing the distribution of
Einstein diameter crossing times in days,
for the same stars. The values for VB~10 and PMI18537+4729
are again marked with green and red lines.
} 
\end{figure}

\end{document}